\documentclass[final,5p,times,twocolumn,authoryear]{elsarticle}

\usepackage{amssymb}
\usepackage{amsmath}
\usepackage{booktabs}
\usepackage{xcolor}
\usepackage{graphicx}
\usepackage{listings}
\usepackage{hyperref}
\hypersetup{
    colorlinks=true,
    linkcolor=blue,
    citecolor=blue,
    urlcolor=blue
}

\journal{New Astronomy}

\begin{document}

\begin{frontmatter}

\title{SpecLess: A Multi-survey Python Package for Machine Learning Classification of Galaxy Morphology and Activity from Photometry}

\author[first,second]{Uzay Aydin\corref{cor1}\fnref{uzay-orcid}}
\ead{uzayaydiin@gmail.com}

\author[second,third]{Mehmet Tanr{\i}ver\fnref{mehmet-orcid}}

\cortext[cor1]{Corresponding author.}

\fntext[uzay-orcid]{ORCID: \url{https://orcid.org/0009-0004-7216-030X}.}
\fntext[mehmet-orcid]{ORCID: \url{https://orcid.org/0000-0002-3263-9680}.}

\affiliation[first]{
    organization={Department of Astronomy and Space Sciences, Graduate School of Natural and Applied Sciences, Erciyes University},
    city={Kayseri},
    postcode={38039},
    country={T\"urkiye}
}

\affiliation[second]{
    organization={Department of Astronomy and Space Science, Faculty of Science, Erciyes University},
    city={Kayseri},
    postcode={38039},
    country={T\"urkiye}
}

\affiliation[third]{
    organization={Erciyes University, Astronomy and Space Science Observatory Application and Research Center},
    city={Kayseri},
    postcode={38039},
    country={T\"urkiye}
}

\begin{abstract}
Galaxy morphology and spectroscopic activity are fundamental 
observational diagnostics of galaxy formation and evolution. 
However, the extent to which these properties can be recovered 
from broadband photometry and preserved across different sky 
surveys remains an important question. In this study, we use 
photometric data from KiDS+VIKING, the DESI Legacy Imaging 
Surveys and SDSS+AllWISE to systematically investigate the 
complementary information provided by optical, near-infrared 
and mid-infrared wavelengths for galaxy classification. 
Our machine learning models achieve macro F1 scores of 
0.9237 and 0.7398 for morphological and activity classification, 
respectively. We show that the contribution of near-infrared 
photometry varies across galaxy populations, yielding an 
improvement of 9.25 percentage points in the classification 
accuracy of red spiral galaxies. Controlled experiments on 
5,235 galaxies common to KiDS+VIKING and DESI demonstrate 
that mid-infrared photometry provides greater complementary 
information for discriminating spectroscopic activity classes 
than additional optical measurements. Furthermore, cross-survey 
transfer experiments conducted on identical galaxies with 
fixed reference labels reveal direction-dependent performance 
losses in the generalizability of photometric classifiers. 
These findings demonstrate that the limitations of photometric 
galaxy classification are governed not only by the choice of 
algorithm but also by the physical properties of galaxy 
populations, wavelength coverage and catalogue-specific 
observational differences. As an outcome of this study, we 
present SpecLess, an open-source Python package providing 
survey-specific classifiers that require no spectroscopic 
information at inference.
\end{abstract}


\end{frontmatter}

%% ======================================================================

\section{Introduction}

Galaxy morphology and spectroscopic activity provide complementary
information on the formation and evolution of galaxies. Morphology
reflects the structural and dynamical history of a galaxy, while
its emission line spectrum provides insight into the physical
processes responsible for the ionisation of the interstellar
medium, including star formation and active galactic nuclei (AGN)
\citep{Conselice2014,HeckmanBest2014,Kewley2006}.
Identifying these properties across large galaxy populations is
therefore important for understanding the relationship between
galactic structure, star formation and nuclear activity.

The optical colours of galaxies are closely related to their
stellar populations and vary systematically with morphological
type \citep{Strateva2001,Baldry2004}. However, colour and morphology
do not necessarily evolve together, as illustrated by the existence
of red spiral and blue elliptical galaxies
\citep{Bamford2009,Schawinski2014}. In spiral galaxies, the
interpretation of optical colours is further complicated by
internal dust attenuation and disc inclination \citep{Masters2010}.
Observations at longer wavelengths can help constrain these
effects. Emission in the near infrared (NIR) traces stellar
populations with reduced sensitivity to dust attenuation, whereas
emission in the mid infrared (MIR) carries signatures of dust
heating associated with star formation and nuclear activity
\citep{Conroy2013,Jarrett2011}. The sensitivity of MIR colours
to these physical processes has led to the development of
photometric diagnostics for identifying AGN candidates
\citep{Stern2012,Mateos2012}.

The photometric and spectroscopic observations provided by the
Sloan Digital Sky Survey (SDSS; \citealt{York2000}), together with
the visual morphological classifications from Galaxy Zoo
\citep{Lintott2008,Lintott2011}, have enabled the construction of
large galaxy samples. Observations from the Wide-field Infrared
Survey Explorer (WISE; \citealt{Wright2010}) have extended the
available photometry to infrared wavelengths. Large surveys such
as the Legacy Survey of Space and Time (LSST) at the Vera C. Rubin
Observatory \citep{Ivezic2019} and Euclid \citep{EuclidMellier2025}
are substantially increasing the number of galaxies with
photometric measurements. Obtaining spectroscopic observations
or detailed visual classifications for every galaxy in such
samples is impractical, motivating the development of automated
classification methods applicable to large photometric catalogues.

Machine learning has been widely applied to galaxy classification
using both images and catalogue photometry. \citet{Ball2004}
demonstrated that morphological and spectral properties of
galaxies could be predicted from photometric parameters.
\citet{Banerji2010} and \citet{Gauci2010} investigated the recovery
of Galaxy Zoo morphological classifications using machine learning
methods. Subsequent studies demonstrated that deep learning can
provide more detailed morphological classifications directly from
galaxy images
\citep{Dieleman2015,DominguezSanchez2018,Walmsley2020,Walmsley2022}.
The applications of deep learning to galaxy surveys have been
extensively reviewed by \citet{HuertasCompanyLanusse2023}, while
software such as Zoobot has made morphological classification
models available for reuse \citep{Walmsley2023}.

In the context of galactic activity, \citet{Cavuoti2014}
investigated the classification of emission line galaxies and
AGN populations using optical photometry, while \citet{Chang2021}
examined the identification of AGN host galaxies by combining
observations from Hyper Suprime-Cam and WISE.
\citet{Daoutis2023} demonstrated that a combination of optical
and infrared colours can distinguish between several
spectroscopically defined galaxy populations.
\citet{Zeraatgari2024} investigated the photometric classification
of galaxies with different emission line properties.
Related methods have also been applied to the separation of stars,
galaxies and quasars, illustrating the applicability of
photometric classification to large astronomical catalogues
\citep{Kurcz2016,Clarke2020}. However, the interpretation of
these classifications depends on the physical information
contained within the available wavelength range and the
definitions of the reference classes.

Differences in filter transmission, observational depth and
photometric measurement procedures can affect the photometric
representation of the same galaxy across surveys. Dependence
on the observational characteristics of the training sample
has also been encountered in photometric redshift estimation
and photometric quasar selection
\citep{Salvato2019,Nakoneczny2021}. The transferability of
morphological classifiers between surveys and their adaptation
to different data domains have previously been investigated
using galaxy images
\citep{DominguezSanchez2019,Ciprijanovic2021,Ciprijanovic2023}.
Similar difficulties in generalisation arise across distinct
galaxy populations. \citet{Aydin2026} reported a substantial
decrease in photometric redshift accuracy when a model trained
on Seyfert 2 galaxies was applied to Seyfert 1 galaxies.
The complementary wavelength coverage of the Kilo-Degree Survey
and VISTA Kilo-Degree Infrared Galaxy Survey (KiDS+VIKING;
\citealt{Wright2024}) and the DESI Legacy Imaging Surveys
\citep{Dey2019} provides an opportunity to investigate these
issues using catalogue photometry and to examine the relative
contributions of optical, NIR and MIR observations.

In this work, we conduct a comprehensive investigation into 
the photometric classification of galaxy morphology and activity 
by exploiting the complementary wavelength coverage of 
KiDS+VIKING and the DESI Legacy Imaging Surveys. To this end, 
we develop two classification models: one designed to distinguish 
between elliptical and spiral galaxies, and another to identify 
star-forming, starburst, Seyfert and LINER galaxies. We investigate 
the role of near-infrared photometry in mitigating the overlap 
between colour and morphology, the contribution of mid-infrared 
photometry to the separation of different spectroscopic activity 
classes, and the extent to which these relationships are preserved 
across photometric catalogues. Using observations of the same 
galaxies from different surveys, we evaluate cross-survey 
transferability under controlled conditions and examine the 
resulting classifications in the context of the physical and 
structural properties of galaxies. Accordingly, we develop 
SpecLess, an open-source Python package designed to characterize 
the morphological and activity properties of large galaxy 
populations using photometric data from wide-field sky surveys. 
By integrating both classification models within a common 
software framework that accommodates survey-specific photometric 
features, we enable the study of galaxy morphology and activity 
through a unified approach.

\section{Data}
\label{sec:data}

\subsection{Morphological sample}
\label{subsec:model1_data}

The morphological reference sample for Model~1 is drawn from 
the debiased visual classifications of Galaxy Zoo~1 
\citep{Lintott2008,Lintott2011}. We select elliptical and spiral 
galaxies with $p_{\rm elliptical}\geq0.80$ and 
$p_{\rm spiral}\geq0.80$, respectively, requiring an SDSS 
spectroscopic classification of \texttt{GALAXY}.

The SDSS sample consists of Galactic extinction corrected 
$ugriz$ magnitudes supplemented by AllWISE $W1$ and $W2$ 
photometry. We exclude invalid optical measurements, require 
\texttt{photometry\_clean=1}, and retain AllWISE counterparts 
within 3~arcsec with photometric quality flags of A or B and 
no contamination in either infrared band. The infrared 
magnitudes are converted from the Vega to the AB system using 
offsets of 2.699 and 3.339~mag, with Galactic extinction 
corrections of $A_{W1}=0.09A_r$ and $A_{W2}=0.05A_r$,
respectively \citep{Davenport2014}. The original selection included a restriction
on optical magnitude uncertainties whose numerical threshold 
was not retained in the archived records. The final sample 
contains 177,766 galaxies (Table~\ref{tab:model1_partition}), 
with seven magnitudes and 21 pairwise colours providing 
28 input features.

\begin{table}[t]
\centering
\caption{Morphological class distribution in the SDSS+AllWISE sample.}
\label{tab:model1_partition}
\footnotesize
\begin{tabular}{lrrr}
\toprule
Class & Train & Test & Total \\
\midrule
Elliptical & 51,553 & 12,888 & 64,441 \\
Spiral     & 90,659 & 22,666 & 113,325 \\
\midrule
Total      & 142,212 & 35,554 & 177,766 \\
\bottomrule
\end{tabular}
\end{table}

We additionally use the fifth data release of the Kilo-Degree 
Survey (KiDS DR5), which incorporates near infrared observations 
from the VISTA Kilo-Degree Infrared Galaxy Survey 
(VIKING; \citealt{Wright2024}). The Galaxy Zoo reference sample 
is matched to KiDS within approximately 1~arcsec, retaining 
sources with valid Gaussian aperture and PSF (GAaP) magnitudes, 
uncertainties and photometric quality flags. We use the 
$u,g,r,i_2$ bands from KiDS and $Z,Y,J,H,K_s$ from VIKING,
hereafter denoting $i_2$ as $i$.
The nine magnitudes and their 36 pairwise colours provide 
45 input features.

The final KiDS+VIKING sample comprises 16,342 unique galaxies, 
including 6,017 ellipticals and 10,325 spirals. Of these, 
12,301 galaxies are common to the SDSS morphology sample, 
while the remaining 4,041 galaxies, comprising 1,298 ellipticals 
and 2,743 spirals, provide an additional evaluation sample. 
Neither morphological probabilities nor structural parameters 
are included among the classification predictors.

\subsection{Spectroscopic sample}
\label{subsec:model2_data}

The reference sample for Model~2 comprises star-forming, 
starburst, Seyfert and low-ionisation nuclear emission line 
region (LINER) galaxies selected from SDSS spectroscopy. 
The classification is based on the Baldwin--Phillips--Terlevich
(BPT) optical diagnostic diagrams
\citep{Baldwin1981,VeilleuxOsterbrock1987}, using
$\log([\mathrm{O\,III}]\lambda5007/\mathrm{H}\beta)$ against 
$\log([\mathrm{N\,II}]\lambda6584/\mathrm{H}\alpha)$ or 
$\log(([\mathrm{S\,II}]\lambda6717+
[\mathrm{S\,II}]\lambda6731)/\mathrm{H}\alpha)$. We adopt the 
demarcations of \citet{Kauffmann2003} and \citet{Kewley2001} 
to distinguish excitation associated with star formation and 
AGN activity, and the [S\,II] diagnostic boundary of 
\citet{Kewley2006} to separate Seyferts and LINERs. Multiple 
spectroscopic observations associated with the same 
\texttt{bestObjID} are resolved using \texttt{snMedian}, 
followed by \texttt{zErr} and \texttt{specObjID} in the event 
of a tie. Objects assigned to more than one reference class 
are excluded before photometric matching.

Star-forming galaxies are selected from the SDSS 
\texttt{STARFORMING} subclass, requiring consistency with 
the adopted star formation criteria in both diagnostic 
diagrams, $S/N\geq5$ in all six emission lines and 
$3<\mathrm{EW}(\mathrm{H}\alpha)<50$~\AA. A random sample 
of 60,000 eligible galaxies is selected using a fixed random 
seed of 42. Starburst galaxies are drawn from the SDSS 
\texttt{STARBURST} subclass, defined in the spectroscopic 
catalogue by star formation with 
$\mathrm{EW}(\mathrm{H}\alpha)>50$~\AA.\footnote{
\url{https://www.sdss3.org/dr8/spectro/catalogs.php}.}
Of the 63,899 available candidates, 60,000 are retained
using an SQL row limit to match the size of the star-forming
sample. The original query ordering was not recorded.

The Portsmouth emission line measurements confirm that all 
60,000 starburst galaxies satisfy 
$\mathrm{EW}(\mathrm{H}\alpha)>50$~\AA. In the [N\,II] 
diagnostic diagram, 54,153 galaxies (90.3 per cent) lie 
below the \citet{Kauffmann2003} boundary, 5,348 
(8.9 per cent) lie between the empirical and theoretical 
boundaries, and 499 (0.8 per cent) lie above the 
\citet{Kewley2001} boundary. These measurements characterise 
the selected sample and are not imposed as additional 
selection criteria. The starburst class therefore follows 
the SDSS spectroscopic definition rather than an independent 
threshold in star formation rate.

Seyfert and LINER galaxies are selected using the adopted 
AGN excitation criteria in both diagnostic diagrams, with 
$S/N\geq3$ in all six emission lines. The two populations 
are separated following \citet{Kewley2006}, with an 
additional requirement of $\mathrm{EW}(\mathrm{H}\alpha)>3$~\AA\ 
for LINERs. No equivalent width restriction or upper 
redshift limit is imposed on the Seyfert sample. The 
resulting spectroscopic reference catalogue contains 
133,886 entries across the four classes 
(Table~\ref{tab:model2_sample}). A subsequent audit identified two
star-forming entries sharing one SDSS object identifier. Both
entries are excluded from the object-level selection audit
(133,884 entries); the archived training partitions are unchanged.

\subsection{Activity photometry}
\label{subsec:model2_photometry}

The spectroscopic reference sample is matched to SDSS and 
AllWISE photometry, requiring valid $ugriz$ magnitudes with 
uncertainties $\leq0.20$~mag and uncontaminated $W1,W2,W3$ 
measurements of photometric quality A or B. We adopt AB 
magnitude offsets of 2.699, 3.339 and 5.174~mag. The archived
photometric reduction uses Galactic extinction corrections
of $A_{W1}=0.09A_r$, $A_{W2}=0.05A_r$ and $A_{W3}=0.13A_r$
\citep[table~3]{Davenport2014}.
The resulting sample
contains 84,652 galaxies, including 16,931 test objects. 
The eight magnitudes and their 28 pairwise colours provide 
36 input features.

The DESI Legacy Imaging Surveys provide complementary 
optical $grz$ measurements and forced WISE photometry 
\citep{Dey2019}. We retain unique primary counterparts 
within 1~arcsec, requiring positive fluxes and inverse 
variances, $S/N\geq3$ in $g,r,z,W1,W2,W3$, and compliance 
with the adopted optical and infrared mask criteria. 
Fluxes expressed in nanomaggies are converted to AB 
magnitudes using $m=22.5-2.5\log_{10}f$, with uncertainties 
calculated as $\sigma_m=(2.5/\ln10)(\sigma_f/f)$. Galactic 
extinction corrections follow $A_\lambda=R_\lambda E(B-V)$, 
where the adopted coefficients are 3.214, 2.165, 1.211, 
0.184, 0.113 and 0.0241 for the six bands, respectively.

The final DESI sample contains 79,827 galaxies with 
27 input features, comprising six magnitudes, their 
associated uncertainties and 15 pairwise colours. 
The sample is divided into training, validation and 
test sets with approximate fractions of 70, 15 and 
15 per cent, respectively (Table~\ref{tab:model2_sample}). 
The requirement for detections in all six bands, 
particularly $W3$, introduces a photometric selection 
relative to the original spectroscopic population.

\begin{table*}[t]
\centering
\caption{Spectroscopic reference sample and the distribution 
of activity classes following DESI photometric selection.}
\label{tab:model2_sample}
\footnotesize
\begin{tabular}{@{}lrrrrr@{}}
\toprule
Class & Spectroscopic reference & DESI total & Train & 
Validation & Test \\
\midrule
Star-forming & 60,000 & 35,875 & 25,112 & 5,381 & 5,382 \\
Starburst    & 60,000 & 34,406 & 24,084 & 5,161 & 5,161 \\
Seyfert      & 7,264  & 5,732  & 4,012  & 860   & 860 \\
LINER        & 6,622  & 3,814  & 2,670  & 572   & 572 \\
\midrule
Total        & 133,886 & 79,827 & 55,878 & 11,974 & 11,975 \\
\bottomrule
\end{tabular}
\end{table*}

\subsection{Matched galaxy samples}
\label{subsec:matched}

To investigate the dependence of classification on survey 
photometry, we construct matched samples with identical 
galaxy identifiers and reference classifications. For 
Model~1, the common SDSS and KiDS sample contains 12,301 
galaxies, comprising 4,719 ellipticals and 7,582 spirals. 
The comparison is restricted to the $u,g,r,i$ magnitudes 
and their six pairwise colours. Each of the 20 stratified 
partitions contains 9,840 training and 2,461 test galaxies, 
with identical sample membership in both catalogues.

For Model~2, the common SDSS and DESI sample comprises 
79,827 galaxies. The original training pool is reduced 
to 10,000 star-forming galaxies, 10,000 starbursts, 
4,012 Seyferts and 2,670 LINERs, while the validation 
and test samples remain unchanged. The resulting sample 
contains 50,631 galaxies across the three partitions. 
The comparison uses the six common magnitudes 
$g,r,z,W1,W2,W3$ and their 15 pairwise colours, excluding 
magnitude uncertainties.

To investigate the complementary wavelength coverage 
of KiDS+VIKING and DESI, we construct a further sample 
of 5,235 spectroscopically classified galaxies with 
photometry in both surveys. The sample contains the 
nine KiDS+VIKING bands and six DESI bands, allowing 
comparisons between optical, near infrared and mid 
infrared photometric configurations for the same 
galaxy population. Measurements from corresponding 
bands are retained separately when the catalogues 
are combined. Identical reference classifications 
and sample partitions are adopted throughout these 
experiments (Table~\ref{tab:kids_desi_activity}).

\begin{table}[t]
\centering
\caption{Activity class distribution in the matched 
KiDS+VIKING and DESI sample.}
\label{tab:kids_desi_activity}
\footnotesize
\setlength{\tabcolsep}{4pt}
\begin{tabular}{lrrrr}
\toprule
Class & Train & Val. & Test & Total \\
\midrule
Star-forming & 1,648 & 352 & 367 & 2,367 \\
Starburst    & 1,547 & 323 & 310 & 2,180 \\
Seyfert      & 306   & 61  & 56  & 423 \\
LINER        & 187   & 44  & 34  & 265 \\
\midrule
Total        & 3,688 & 780 & 767 & 5,235 \\
\bottomrule
\end{tabular}
\end{table}

\section{Methods}
\label{sec:experiments}

\subsection{Classification algorithms}
\label{subsec:classifiers}

We employ logistic regression (LR), $k$ nearest neighbours 
(KNN) and extreme gradient boosting (XGBoost; 
\citealt{ChenGuestrin2016}) to investigate morphological and 
activity classification from broadband photometry.

For Model~1, the input features are standardised for LR and 
KNN. Logistic regression uses balanced class weights and a 
maximum of 3,000 iterations, while KNN adopts $k=15$, 
Euclidean distance and distance weighting. The XGBoost 
classifier used in the SDSS and KiDS comparison adopts 
350 estimators, a maximum tree depth of 5, a learning rate 
of 0.05 and sampling fractions of 0.8 for both observations 
and features. The separate pretrained SDSS morphology model
uses 300 estimators, maximum depth 6 and learning rate 0.1.

For Model~2, the XGBoost classifier uses 300 estimators, 
a maximum tree depth of 5, a learning rate of 0.05 and 
sampling fractions of 0.85. We adopt the 
\texttt{multi:softprob} objective and multiclass logarithmic 
loss as the evaluation metric. Random seed 42 is used 
where applicable.

\subsection{Morphological classification}
\label{subsec:morphology_method}

The dependence of morphological classification on the 
photometric catalogue is investigated using the 12,301 
galaxies common to SDSS and KiDS. The classifiers are 
trained independently on the $u,g,r,i$ magnitudes and 
their six pairwise colours and evaluated over 20 
identical stratified partitions. Models trained in each 
survey are evaluated using photometry from both catalogues. 
We repeat the XGBoost experiments using only the six 
colours to assess the contribution of apparent magnitudes.

To isolate the contribution of NIR photometry, we train
unweighted LR models ($C=1$, 2,000 iterations) on the
original 12,301 KiDS galaxies and evaluate them on the
4,041 additional objects, using either $ugriZ$ or
all nine magnitudes. The shorter wavelength set includes
the VIKING $Z$ band. We repeat the comparison with
histogram gradient boosting (150 iterations, learning rate
0.05, at most 15 leaves per tree, minimum 30 samples per
leaf and $L_2$ regularisation of 1). Red spirals are defined
by $u-r\geq2.7706$, the lower quartile of the elliptical
colour distribution in the training sample, and blue
ellipticals by $u-r\leq2.5861$, the upper quartile for spirals.
These thresholds are fixed before evaluation. We examine
the dependence of recovery on the propagated $J-K_s$
uncertainty, projected axis ratio and Galaxy Zoo vote
fraction. For red spirals initially misclassified by the
optical model, a logistic regression estimates the association
between NIR recovery and uncertainty, controlling for $u-r$,
$r$ and angular half-light radius.

\subsection{Activity classification}
\label{subsec:activity_method}

The DESI activity classifier is evaluated using the 
27 photometric features and fixed sample partition 
described in Section~\ref{subsec:model2_photometry}. 
We examine the effects of class weighting, feature 
selection and probability calibration, with dummy and 
logistic regression classifiers providing reference 
performance. An additional 20 stratified partitions 
are used to assess sensitivity to sample partitioning.

The 5,235 galaxies common to KiDS+VIKING and DESI are
used to compare wavelength coverage. We fit balanced LR
models to standardised magnitudes, using up to 3,000 iterations
and the partitions in Table~\ref{tab:kids_desi_activity}.
The optical, NIR and MIR combinations are explored on the
same test sample; selected comparisons are repeated on the
validation sample. We also examine the consistency of
Seyfert and LINER labels with the \citet{Kewley2006}
[S\,II] demarcation in 598 development objects with
$S/N\geq3$ in the diagnostic lines.

Spectroscopic measurements are used only for the subsequent
analysis of classification errors. We measure starburst recall
as a function of H$\alpha$ equivalent width in the fixed DESI
test set. A separate archive contains predictions from three
algorithms for 11,974 galaxies; its membership could not be
verified against the original validation partition and is
therefore analysed separately. We compare the NIR colours
of consistently misclassified and correctly classified
starbursts using regression with optical colours, redshift
and $r$ as covariates. For Seyferts, we estimate the odds
of assignment to LINER rather than Seyfert as a function of
logarithmic H$\alpha$ equivalent width and BPT line ratios.

For the matched Seyfert sample, ridge regression trained
on 306 objects is used to remove the dependence of MIR
colour and excitation on redshift, KiDS photometry and the
adopted spectroscopic covariates. Residual correlations are
evaluated on 61 validation objects, and the contribution of
MIR colours to line ratio prediction is assessed on 56 test
objects. For starbursts, we model the difference in $g-r$
between KiDS and DESI using redshift (including its quadratic
term), KiDS colour and $r$ magnitude, then add logarithmic
H$\alpha$ and [O\,III] equivalent widths and their interactions
with redshift. We evaluate the reduction in mean absolute error
on 310 test galaxies and compare it with 300 permutations of
the line measurements within training strata of redshift and
colour. These are exploratory physical diagnostics.

\subsection{Transferability between surveys}
\label{subsec:design}

For activity classification, models are trained 
independently using the common SDSS and DESI 
feature set, comprising the $g,r,z,W1,W2,W3$
magnitudes and their 15 pairwise colours. Balanced class
weights are used in the matched SDSS and DESI fits. The
experiments are repeated using only the colours
and separately retrained 21-feature controls with the same
training identities to assess the contribution of apparent
magnitudes to survey transferability.

For the morphology transfer fits, we estimate the median
magnitude difference between the catalogues from the training
objects in each band. We subtract this offset from the target
photometry to approximate the source magnitude system and
recalculate the corresponding colours.

To distinguish the effects of optical and mid infrared 
photometry on activity classification, we independently 
exchange the $grz$ and $W1,W2,W3$ measurements between 
the SDSS and DESI representations of the same test 
galaxies. All colours are recalculated following each 
substitution, while the fitted classifiers remain 
unchanged.

The dependence of transferability on training sample 
composition is examined using 20 alternative 
realisations of the activity training population. 
Each contains 10,000 star-forming galaxies, 10,000 
starbursts and all available Seyfert and LINER galaxies 
from the original training pool. The validation and 
test populations remain fixed.

\subsection{Performance metrics and statistical analysis}
\label{subsec:metrics}

Classification performance is assessed using accuracy, 
balanced accuracy, macro F1 and ROC AUC 
where probability estimates are available. Precision, 
recall and normalised confusion matrices are used to 
examine the performance of individual galaxy classes. 
Macro F1 gives equal weight to the four activity classes.
For a model trained on catalogue $A$ and evaluated on $B$,
$F_{A\rightarrow B}$ denotes its macro F1. We distinguish
the loss relative to its original performance,
$D_{\mathrm{src}}=F_{A\rightarrow A}-F_{A\rightarrow B}$,
from the deficit relative to a classifier trained on the
target catalogue,
$D_{\mathrm{tgt}}=F_{B\rightarrow B}-F_{A\rightarrow B}$.
Only the latter holds the test photometry fixed.

Standard deviations across repeated partitions and training
realisations describe variation rather than uncertainty from
independent samples. We obtain paired 95 per cent bootstrap
intervals from 2,000 stratified resamples for the SDSS and DESI
transfer contrasts. The subsequent KiDS+VIKING analyses use
paired bootstrap resampling for differences in predictions
and regression errors. These intervals are conditional on
the fitted models and available evaluation objects.

The probability calibration of the native DESI activity
classifier is assessed using the expected calibration
error (ECE) for the maximum predicted probability, with
15 confidence bins and 500 bootstrap realisations.

%% ======================================================================

\section{Results}
\label{sec:results}

The performance of the four survey-specific XGBoost models is
summarised in Table~\ref{tab:packaged}. Their scores describe
distinct feature sets and samples and should not be interpreted
as matched comparisons between surveys.

\begin{table*}[t]
\centering
\caption{Performance of the four pretrained XGBoost classifiers on
their respective test samples. For the DESI activity model,
ROC AUC is the macro-averaged one-versus-rest area under the curve.}
\label{tab:packaged}
\footnotesize
\begin{tabular}{@{}llrrrrr@{}}
\toprule
Task & Survey & $N_{\rm test}$ & Accuracy & Bal. accuracy & Macro F1 & ROC AUC \\
\midrule
Morphology & SDSS+AllWISE & 35,554 & 0.9375 & 0.9388 & 0.9332 & 0.9826 \\
Morphology & KiDS+VIKING & 3,269 & 0.9284 & 0.9274 & 0.9237 & 0.9775 \\
Activity & SDSS+AllWISE & 16,931 & 0.8217 & 0.7865 & 0.7637 & 0.9572 \\
Activity & DESI Legacy & 11,975 & 0.7986 & 0.7255 & 0.7398 & 0.9476 \\
\bottomrule
\end{tabular}
\end{table*}

\subsection{Morphological classification}
\label{subsec:results_model1}

The KiDS+VIKING classifier achieves a macro F1 of 0.9237
without structural measurements among its predictors.
In the separate evaluation of 4,041 additional galaxies,
including the four longer wavelength VIKING bands $YJHK_s$
alongside $ugriZ$ increases the macro F1 score from
0.9401 to 0.9481 for logistic regression. However, the
improvement is not uniform across the morphological
populations (Table~\ref{tab:morphology_nir}).
The largest increase is observed for red spirals, whose
classification accuracy rises from 64.04 to 73.29 per cent,
corresponding to an improvement of 9.25 percentage points
(95 per cent confidence interval: $[5.14,13.36]$).
A comparable improvement of 8.22 percentage points is
obtained using histogram gradient boosting, indicating
that the result is not restricted to a single classification
algorithm.

\begin{table}[t]
\centering
\caption{Accuracy of logistic regression on the 4,041 additional
KiDS+VIKING galaxies with and without NIR photometry.}
\label{tab:morphology_nir}
\footnotesize
\begin{tabular}{lrrr}
\toprule
Population & $N$ & Optical & Optical+NIR \\
\midrule
Red spiral       & 292   & 0.6404 & 0.7329 \\
Other spiral     & 2,451 & 0.9894 & 0.9874 \\
Blue elliptical  & 50    & 0.3400 & 0.4000 \\
Other elliptical & 1,248 & 0.9607 & 0.9639 \\
\bottomrule
\end{tabular}
\end{table}

For blue ellipticals, the accuracy increases from 34 to
40 per cent, although the confidence interval includes zero.
The negligible changes observed for the remaining galaxies
suggest that the additional discriminatory information
provided by the NIR bands is particularly relevant to
red spirals within this sample.

The contribution of NIR photometry also depends on
measurement quality. Dividing the 292 red spirals at the
median propagated $J-K_s$ colour uncertainty of 0.0035~mag
reveals an accuracy improvement of 17.12 percentage points
in the lower uncertainty group, compared with 1.37 percentage
points in the higher uncertainty group. Among the galaxies
initially misclassified using optical photometry, the odds
of correct classification after including the NIR bands are
higher in the lower uncertainty group, with an adjusted odds
ratio of 3.31 (95 per cent confidence interval: $[1.04,12.71]$)
after accounting for optical colour, apparent magnitude and
angular half-light radius.

Red spirals have a median projected axis ratio of
$b/a=0.346$, compared with 0.485 for the remaining spirals.
For the 274 red spirals with $b/a\leq10^{-0.2}\simeq0.631$,
the inclusion of NIR photometry increases the number of
correct classifications from 182 to 211 (66.42 to 77.01
per cent). Nevertheless, the recovery rates
among initially misclassified galaxies do not differ
significantly between red and other spirals within this
orientation range. The improvement also persists among
152 galaxies with Galaxy Zoo spiral probabilities
$p_{\rm spiral}\geq0.90$, for which the accuracy increases
by 13.16 percentage points (95 per cent interval:
$[7.24,19.08]$).

\subsection{Spectroscopic activity classification}
\label{subsec:results_model2}

The DESI Legacy XGBoost classifier achieves a macro F1 of
0.7398 on 11,975 test galaxies. In comparison,
logistic regression achieves a macro F1 score of 0.7008 and
a balanced accuracy of 0.7496, while the dummy classifier
returns a macro F1 score of 0.1550. The higher balanced
accuracy of logistic regression relative to XGBoost (0.7255)
indicates higher mean recall across classes, whereas its
lower macro F1 reflects a less favourable balance of
precision and recall.

The classification performance varies considerably among
the four spectroscopic classes (Table~\ref{tab:model2_classwise}).
Star-forming and starburst galaxies are recovered more
accurately than Seyfert and LINER galaxies. Confusion occurs
predominantly between star-forming and starburst galaxies,
while Seyferts and LINERs are also frequently assigned to
other activity classes.

\begin{table}[t]
\centering
\caption{Classification performance of the DESI Legacy
XGBoost activity classifier on 11,975 test galaxies.}
\label{tab:model2_classwise}
\footnotesize
\setlength{\tabcolsep}{4pt}
\begin{tabular}{lrrrr}
\toprule
Class & Precision & Recall & F1 & $N$ \\
\midrule
Star-forming & 0.7853 & 0.8400 & 0.8117 & 5,382 \\
Starburst    & 0.8401 & 0.8029 & 0.8211 & 5,161 \\
Seyfert      & 0.6905 & 0.6174 & 0.6519 & 860 \\
LINER        & 0.7112 & 0.6416 & 0.6746 & 572 \\
\bottomrule
\end{tabular}
\end{table}

Across 20 stratified partitions, the mean macro F1 score
is $0.7494\pm0.0061$. The LINER population exhibits the
largest variation among the individual class F1 scores.
Introducing balanced class weights increases Seyfert and
LINER recall to 0.6977 and 0.7570, respectively, although
the overall macro F1 score decreases to 0.7280.

The contribution of different photometric measurements
is examined through feature ablation. Restricting the DESI
classifier to optical or MIR photometry reduces the macro
F1 score to 0.6000 and 0.6909, respectively. Removing the
$W3$ band reduces the score to 0.7053, whereas excluding
magnitude uncertainties produces only a minor decrease.
A model using only the $g-r$, $W1-W2$ and $W2-W3$ colours
achieves a macro F1 score of 0.7035, indicating that these
colours retain a substantial fraction of the information
available to the full 27 feature classifier.

The mean maximum predicted probability of the DESI
classifier is 0.8064, compared with an empirical accuracy
of 0.7986. The expected calibration error is 0.0153
(95 per cent confidence interval: $[0.0108,0.0227]$),
indicating a small degree of overconfidence within the
DESI test sample.

\subsection{Photometric information and survey transferability}
\label{subsec:results_transfer}

Table~\ref{tab:model1_paired} presents the results of the
controlled morphology experiments using the same 12,301
galaxies and common optical features in SDSS and KiDS.
The XGBoost classifier achieves comparable performance
when trained and evaluated within each survey, with macro
F1 scores of 0.9049 and 0.9065 for SDSS and KiDS,
respectively. Applying the trained models to the alternative
photometric catalogue reduces their performance, with a
larger decrease observed for the transfer from KiDS to SDSS.

\begin{table*}[t]
\centering
\caption{Macro F1 scores for the morphological classification
of the same galaxies using SDSS and KiDS photometry.
Values are the mean and standard deviation across 20
stratified partitions.}
\label{tab:model1_paired}
\footnotesize
\begin{tabular}{lccc}
\toprule
Training and test catalogues &
Logistic regression & $k$-NN & XGBoost \\
\midrule
SDSS $\rightarrow$ SDSS
& $0.8867\pm0.0060$
& $0.8996\pm0.0068$
& $0.9049\pm0.0057$ \\
SDSS $\rightarrow$ KiDS
& $0.8797\pm0.0075$
& $0.8937\pm0.0057$
& $0.8912\pm0.0061$ \\
KiDS $\rightarrow$ KiDS
& $0.8922\pm0.0057$
& $0.9032\pm0.0057$
& $0.9065\pm0.0044$ \\
KiDS $\rightarrow$ SDSS
& $0.8622\pm0.0054$
& $0.8495\pm0.0093$
& $0.8695\pm0.0094$ \\
\bottomrule
\end{tabular}
\end{table*}

The photometric differences between SDSS and KiDS depend
on morphological type. The median difference in $r$-band
magnitude, defined as KiDS minus SDSS, is 0.7092~mag for
ellipticals and 0.9723~mag for spirals, whereas the median
$g-r$ colour differences are 0.0182 and 0.0550~mag,
respectively. Applying additive corrections derived from
the training sample increases the transfer macro F1 scores
from 0.8912 to 0.8994 for SDSS to KiDS and from 0.8695
to 0.8778 in the opposite direction. Restricting the models
to optical colours reduces the KiDS to SDSS target deficit
from 0.0354 to 0.0245, while that for SDSS to KiDS changes
from 0.0153 to 0.0147.

\begin{figure*}[t]
\centering
\includegraphics[width=\textwidth]
{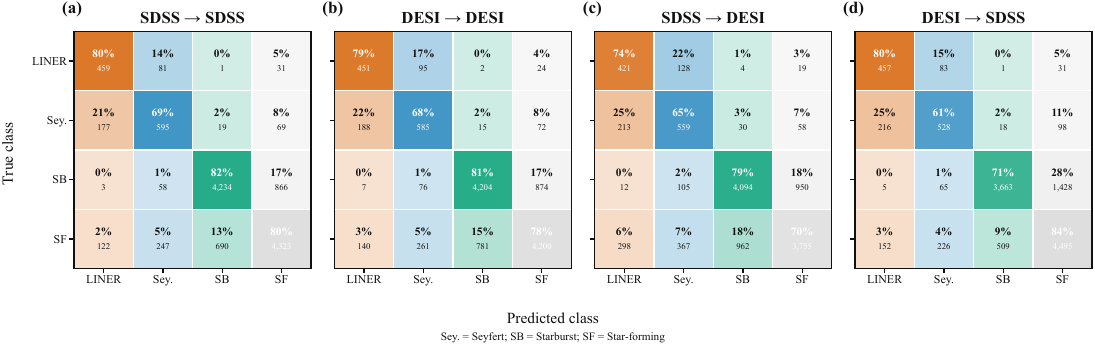}
\caption{Row-normalised confusion matrices for the matched
activity experiments: (a) SDSS to SDSS, (b) DESI to DESI,
(c) SDSS to DESI and (d) DESI to SDSS.
The same 11,975 test galaxies and reference labels are used
in each panel. SF and SB denote star-forming and starburst
galaxies, respectively.}
\label{fig:model2_crosssurvey_confusion}
\end{figure*}

A similar dependence is observed for activity classification
(Fig.~\ref{fig:model2_crosssurvey_confusion}). On the same
DESI test galaxies, the SDSS and DESI models attain macro F1
scores of 0.6617 and 0.7277, respectively. On SDSS photometry,
the corresponding scores are 0.7458 and 0.7052. Relative to
the target-trained models, the transfer deficits are 0.0660
and 0.0406, with paired 95 per cent intervals of
$[0.0569,0.0757]$ and $[0.0307,0.0501]$, respectively.
The losses relative to the original models' own catalogue
scores are 0.0841 and 0.0225. The directional asymmetry
persists across 20 alternative training selections, with
mean target deficits of 0.0773 and 0.0292, respectively.
Colour-only activity refits reduce the SDSS to DESI target
deficit from 0.0693 to 0.0345 but increase the reverse
deficit from 0.0358 to 0.0868, relative to the corresponding
21-feature controls retrained for this comparison, rather
than the saved models in Fig.~\ref{fig:model2_crosssurvey_confusion}.
The 15 input colours contain only five independent colour
contrasts. Removing apparent magnitudes thus does not
consistently improve transferability.

\begin{figure*}[t]
\centering
\includegraphics[width=\textwidth]
{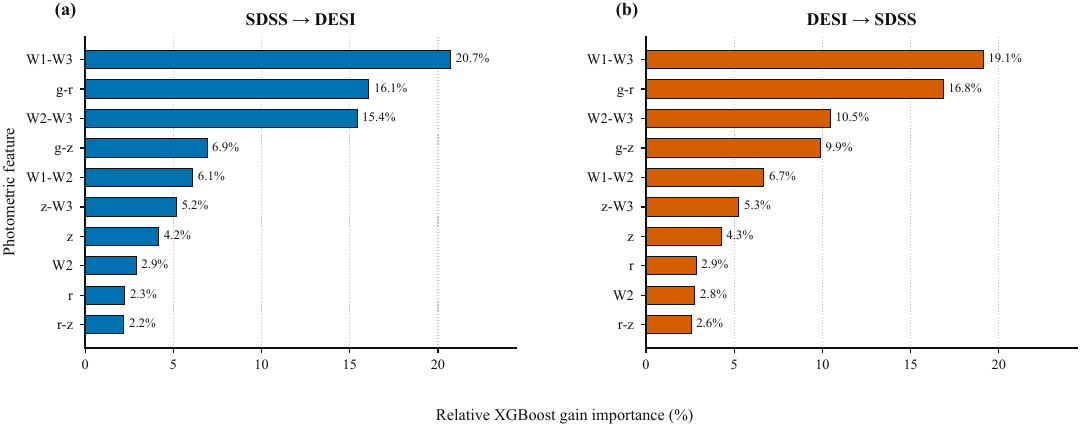}
\caption{XGBoost feature importance (normalised gain) for the
(a) SDSS+AllWISE and (b) DESI activity classifiers trained with the
same 21 photometric predictors on matched galaxies.
The arrows in the panel headings indicate the transfer
direction of each source model, not feature importances
measured from the target photometry.}
\label{fig:model2_crosssurvey_feature_importance}
\end{figure*}

Replacing the optical and MIR measurements independently
reveals that both wavelength regions contribute to the
transfer deficit of the SDSS trained classifier. Replacing
only the optical measurements reduces the macro F1 score
by 0.0590, while replacing only the MIR measurements
reduces it by 0.0540. For the DESI trained classifier,
the corresponding reductions are 0.0040 and 0.0425.
The optical substitution in the DESI model is consistent
with zero change (95 per cent interval: $[-0.0133,0.0057]$
for transferred minus original performance).
The effects of the two substitutions are not additive,
and these mixed photometric inputs should be regarded as
diagnostic experiments. The gain-based feature rankings
(Fig.~\ref{fig:model2_crosssurvey_feature_importance})
identify $W1-W3$, $g-r$ and $W2-W3$ as the three most
important predictors in both matched XGBoost models.
These rankings are conditional on the fitted trees and
correlated input features.

The matched KiDS+VIKING and DESI activity sample provides
a further test of catalogue dependence. Using only the
three corresponding bands ($grZ$ from KiDS+VIKING
and $grz$ from DESI), the two
classifiers achieve macro F1 scores of 0.5164 and 0.4966,
respectively. Transferring the classifiers between surveys
reduces these scores to 0.4683 and 0.4777, respectively,
and changes the predicted classes of 18.38 and 23.08 per
cent of the same 767 test galaxies.

We next examine the contribution of additional wavelength
coverage using the same galaxy sample and identical
training and test partitions
(Table~\ref{tab:multisurvey_photometry}).
Extending $ugriZ$ with the four longer wavelength VIKING
bands ($YJHK_s$) improves the macro F1 score from
0.6268 to 0.6695, corresponding to an increase of 0.0427
(95 per cent confidence interval: $[0.0156,0.0693]$).
Seyfert recall increases from 42.86 to 53.57 per cent,
whereas LINER recall decreases from 88.24 to 82.35 per cent.

\begin{table*}[t]
\centering
\caption{Activity classification using different photometric
combinations on the matched KiDS+VIKING and DESI sample.
All test and validation scores refer to the same 767 and
780 galaxies, respectively.
Differences quoted in the text are calculated before rounding.}
\label{tab:multisurvey_photometry}
\footnotesize
\begin{tabular}{lrrr}
\toprule
Photometric measurements & $N_{\rm phot}$ &
Test macro F1 & Validation macro F1 \\
\midrule
KiDS+VIKING $grZ$                  & 3  & 0.5164 & 0.5366 \\
KiDS+VIKING $ugriZ$                & 5  & 0.6268 & 0.6164 \\
KiDS+VIKING                        & 9  & 0.6695 & 0.6950 \\
DESI $grz$                         & 3  & 0.4966 & 0.5221 \\
DESI $grzW1W2W3$                   & 6  & 0.7137 & 0.6820 \\
KiDS+VIKING + DESI optical         & 12 & 0.6828 & 0.6863 \\
DESI + KiDS+VIKING $ugriZ$         & 11 & 0.7477 & 0.7162 \\
KiDS+VIKING + $W1W2$               & 11 & 0.7121 & 0.7262 \\
KiDS+VIKING + $W1W2W3$             & 12 & 0.7648 & 0.7251 \\
KiDS+VIKING + DESI                 & 15 & 0.7571 & 0.7259 \\
\bottomrule
\end{tabular}
\end{table*}

Combining all 15 photometric measurements from the two
catalogues increases the macro F1 score to 0.7571.
To distinguish the contribution of additional wavelength
coverage from that of repeated optical measurements,
we separately add the DESI optical and MIR photometry
to the nine KiDS+VIKING bands. The optical measurements
increase the macro F1 score by 0.0134, whereas the MIR
measurements produce an increase of 0.0954.
The difference between these improvements is 0.0820
(95 per cent confidence interval: $[0.0345,0.1275]$).

The combination of KiDS+VIKING and the three MIR bands
achieves the highest test score of 0.7648, exceeding DESI
alone by 0.0511 (95 per cent interval: $[0.0120,0.0891]$).
The difference relative to all 15 measurements is not
statistically conclusive. On the validation sample, the
same combination achieves 0.7251, exceeding the DESI
score of 0.6820 by 0.0431 ($[0.0078,0.0816]$); its
advantage over KiDS+VIKING alone is inconclusive.

Adding $W3$ to KiDS+VIKING and $W1,W2$ increases test
macro F1 by 0.0527 (95 per cent interval: $[0.0224,0.0874]$),
but changes validation macro F1 by $-0.0011$
($[-0.0346,0.0322]$). Without $W3$, the combined model
still exceeds KiDS+VIKING alone on the test sample by 0.0426
($[0.0034,0.0808]$), but not DESI alone. The evidence for
the additional contribution of $W3$ is thus sample dependent.

\subsection{Spectroscopic properties and exploratory photometric diagnostics}
\label{subsec:physical_properties}

All 598 Seyfert and LINER development objects assessed
against the [S\,II] boundary agree with their reference
labels. This is a consistency test of labels derived from
spectroscopic diagnostics, not an independent validation.
The recovery of the spectroscopic starburst class is
strongly associated with H$\alpha$ equivalent width.
Among the 5,161 starburst galaxies in the DESI test
sample, the classification recall increases from 0.6132
at $50\leq\mathrm{EW}(\mathrm{H}\alpha)<60$~\AA\
to 0.9925 at $\mathrm{EW}(\mathrm{H}\alpha)\geq150$~\AA\
(Table~\ref{tab:model2_ew_starburst}).
The fraction incorrectly classified as star-forming
decreases from 37.6 per cent in the lowest equivalent
width interval to 2.7 per cent at equivalent widths of
100~\AA\ or greater.
Thus, the photometric distinction between the two
catalogue classes becomes less reliable close to the
equivalent width threshold used in their definition.

\begin{table}[t]
\centering
\caption{Recovery of the DESI starburst reference class
as a function of H$\alpha$ equivalent width. The final
column gives the fraction classified as star-forming.}
\label{tab:model2_ew_starburst}
\footnotesize
\begin{tabular}{lrrr}
\toprule
EW(H$\alpha$) (\AA) & $N$ & Recall & SB$\rightarrow$SF \\
\midrule
$50$--$60$   & 1,528 & 0.6132 & 0.3757 \\
$60$--$75$   & 1,342 & 0.7906 & 0.2042 \\
$75$--$100$  & 1,034 & 0.8946 & 0.1015 \\
$100$--$150$ & 722   & 0.9557 & 0.0416 \\
$\geq150$    & 535   & 0.9925 & 0.0075 \\
\bottomrule
\end{tabular}
\end{table}

Additional information is obtained from the NIR colours
of galaxies for which different classification algorithms
produced the same incorrect predictions. Among 761
galaxies with both KiDS+VIKING photometry and archived
DESI predictions, 43 spectroscopic starbursts are
consistently classified as star-forming by logistic
regression, $k$-NN and XGBoost. These galaxies exhibit
a median $J-K_s$ colour of 0.1269~mag, compared with
0.1707~mag for the 220 correctly classified starbursts.
After accounting for optical colours, redshift and
apparent magnitude, the estimated difference is
$-0.0583$~mag (95 per cent confidence interval:
$[-0.0878,-0.0275]$). Nevertheless, the improvement from
including NIR photometry in a separate binary classification
experiment is not statistically conclusive.

The confusion between Seyfert and LINER galaxies is also
associated with their emission line properties.
Among the 11,974 archived XGBoost predictions, the 107 Seyferts
classified as LINERs have a median H$\alpha$ equivalent
width of 7.89~\AA, compared with 14.90~\AA\ for the
519 correctly classified Seyferts. This association
persists after accounting for the
$[\mathrm{O\,III}]/\mathrm{H}\beta$ and
$[\mathrm{N\,II}]/\mathrm{H}\alpha$ line ratios, with an adjusted
odds ratio of 0.249 ($[0.170,0.324]$) for assignment to LINER
rather than Seyfert per standard deviation increase in log
equivalent width.

We further investigate whether MIR photometry contains
additional information about the excitation properties
of Seyfert galaxies. After accounting for redshift,
optical and NIR photometry, the remaining BPT line ratios,
the Balmer decrement and emission line measurement
quality, the residual $W1-W2$ colour is positively
correlated with the residual
$\log([\mathrm{O\,III}]/\mathrm{H}\beta)$ in the
validation sample of 61 Seyferts
($\rho=0.378$, $p=0.0027$).
However, the test sample of 56 Seyferts provides no conclusive
evidence that MIR colours reduce the line ratio prediction
error ($\Delta\mathrm{MAE}=0.0131$~dex, 95 per cent interval:
$[-0.0035,0.0291]$ for the reduction).

Finally, the differences in photometric colours between
KiDS and DESI depend on the spectroscopic activity class.
For starburst galaxies, the difference in $g-r$ colour
between KiDS and DESI (KiDS minus DESI) correlates with redshift
in the training sample ($\rho=0.383$). Including the equivalent widths of
H$\alpha$ and $[\mathrm{O\,III}]$ in the regression of
these colour differences reduces the mean absolute error
from 0.03044 to 0.02937~mag on the test sample, a reduction
of 0.00107~mag (95 per cent interval: $[0.00012,0.00206]$).
None of the 300 permutations yields a reduction as large
as that observed ($p=1/301\simeq0.0033$, the minimum
attainable Monte Carlo $p$-value). The effect is small
in absolute photometric units.

The matched KiDS+VIKING and DESI sample contains only
3.9 per cent of the spectroscopic reference population
after excluding ambiguous identifiers. The retained
fractions are 3.95, 3.63, 5.82 and 4.00 per cent for
star-forming, starburst, Seyfert and LINER galaxies,
respectively. The redshift distributions of the matched
and remaining objects also differ for three of the four
classes. The matched sample is therefore not representative
of the full spectroscopic reference catalogue.

%% ======================================================================
\section{Discussion}
\label{sec:discussion}

\subsection{Wavelength coverage and galaxy populations}
\label{subsec:discussion_morphology}

\citet{Banerji2010} identified red spirals and blue ellipticals
among galaxies for which neural network classifications
disagreed with Galaxy Zoo morphologies. Their classifiers
used SDSS photometry and structural measurements, whereas
our KiDS+VIKING experiment isolates the contribution of
wavelength coverage without structural predictors. The
9.25 percentage point improvement for red spirals
(Table~\ref{tab:morphology_nir}), with negligible changes
for the main spiral and elliptical populations, shows that
the additional information in $YJHK_s$ is most useful where
optical colour and morphology disagree. This extends the
earlier findings by identifying an observational configuration
in which a subset of such misclassifications can be recovered.
It does not establish that integrated photometry can replace
image-based morphology, and the high Galaxy Zoo vote threshold
restricts the comparison to relatively unambiguous reference
morphologies.

\citet{Smethurst2022} found little benefit
from adding NIR bands to optical photometry when morphology was
inferred through single-colour cuts on Galaxy Zoo 2 samples.
Our modest aggregate improvement is consistent with a limited
overall contribution from longer wavelengths. The analysis by
population, however, reveals a measurable benefit for red spirals
when multiband photometry is used jointly. The two experiments
differ in reference labels, sample selection and classification
method; the comparison illustrates how aggregate measures can
obscure gains within specific populations.

The behaviour of red spirals is consistent with the known
partial decoupling of colour and morphology
\citep{Strateva2001,Bamford2009,Masters2010}. NIR
measurements are less sensitive to dust attenuation than
optical measurements and probe a different mixture of
stellar populations \citep{Conroy2013}. Their contribution
to red spiral recovery is associated with photometric
precision, even after accounting for optical colour, apparent
magnitude and angular size. Although the red spirals have
smaller projected axis ratios, the recovery fractions of
initially misclassified red and other spirals do not differ
significantly in the low axis ratio subset. The inclination
result therefore does not establish dust as the cause of
the improvement; intrinsically red disc populations and
measurement effects remain possible contributors. The small
blue elliptical sample prevents an equivalent conclusion
for that population.

For activity classification, \citet{Daoutis2023} showed that
three optical and MIR colours ($g-r$, $W1-W2$ and $W2-W3$)
are effective predictors of five spectroscopically defined
classes. Restricting our DESI classifier to the same colours
yields a macro F1 of 0.7035. This is a comparison of
photometric information content within our sample, not a
reproduction of their random forest or a comparison of
equivalent class definitions. \citet{Zeraatgari2024}
also distinguished star-forming, starburst and AGN spectra
using SDSS and AllWISE photometry in a broader astronomical
source classification scheme. Our analysis addresses a
complementary question: whether VIKING and DESI optical and
MIR measurements provide complementary information for
the same galaxies. Adding the MIR bands to KiDS+VIKING
yields a larger gain than adding the same number of DESI
optical measurements. This is consistent with the differing
contributions of stellar light and dust emission across the
NIR and MIR \citep{Conroy2013,Jarrett2011}. The improvement
over DESI alone persists in validation, but the additional
benefit of $W3$ does not. The complementary wavelength
coverage is thus informative within the selected population,
without establishing an optimal universal band combination.

\subsection{Spectroscopic properties and classification errors}
\label{subsec:discussion_activity}

The distinction between star-forming and starburst galaxies
is particularly sensitive to the reference class definition.
Unlike the five-class diagnostic of \citet{Daoutis2023}, our
scheme separates starbursts from other star-forming galaxies
through the SDSS spectroscopic subclass. The increase in
recall from 61.32 per cent near the lower H$\alpha$ equivalent
width boundary to 99.25 per cent at EW(H$\alpha$)$\geq150$~\AA
illustrates how recovery of discrete labels varies with a
continuous spectroscopic observable. Broadband colours trace
stellar continua and cannot be expected to reproduce a sharp
equivalent width boundary exactly. The result quantifies the
reliability of recovering the adopted subclass, rather than
independently identifying physically defined starbursts.
The adjusted $J-K_s$ difference between consistently
misclassified and correctly classified starbursts suggests
an association with NIR spectral shape, although the separate
binary experiment does not establish a corresponding improvement
in classification.

Classical MIR colour criteria were developed to select AGN
with characteristic infrared emission \citep{Stern2012,Mateos2012};
they do not by themselves reproduce our four spectroscopic
classes. For Seyferts, the association between weaker H$\alpha$
equivalent width and assignment to LINER persists after
conditioning on two optical diagnostic ratios. This is consistent
with the possibility that host galaxy light dilutes the
photometric signatures of nuclear activity, although equivalent
width alone cannot identify the cause. LINER spectra may arise
from several excitation mechanisms
\citep{Ho2008,CidFernandes2011,Belfiore2016}. The exploratory
residual association between MIR colour and Seyfert excitation
is likewise not confirmed as a predictive improvement on the
separate test sample. These results identify limits to the
recovery of spectroscopic labels from broadband photometry;
they do not establish independent diagnostics of the underlying
ionisation mechanism.

\subsection{Survey dependence and limitations}
\label{subsec:discussion_transfer}

\citet{DominguezSanchez2019} investigated the adaptation of
image-based SDSS morphology classifiers to observations from
the Dark Energy Survey. Related domain adaptation studies
address differences between image domains
\citep{Ciprijanovic2021,Ciprijanovic2023}. Our matched catalogue
experiments isolate a distinct aspect of this problem:
transfer deficits occur even without a change in galaxy
identities or reference labels. For morphology, the KiDS
and SDSS magnitude offsets vary with reference class, and
neither a restriction to colours nor additive corrections
fully removes the deficit. The activity experiments further
show that the relative effect of optical and MIR substitution
depends on the direction of transfer. Hence, catalogue
measurements cannot in general be treated as interchangeable
inputs, even when nominally corresponding bands are available.

Differences in aperture definitions, filter transmission,
forced photometry and extinction prescriptions are possible
contributors. For instance, the adopted $W3$ extinction
correction relative to the optical $r$ band differs between
SDSS+AllWISE and DESI (Section~\ref{subsec:model2_photometry}),
and its contribution has not been isolated. The association
between the KiDS minus DESI $g-r$ difference and spectroscopic
properties of starbursts is detectable in the matched sample,
but the reduction in prediction error is only 0.00107~mag.
These exploratory tests motivate further work on physically
informed photometric harmonisation; they do not establish
a unique instrumental or astrophysical cause of the offsets.

The spectroscopic reference classes and photometric selection
limit the interpretation of the results. The morphological
sample favours galaxies with relatively secure visual labels,
while the matched activity population contains a small and
class-dependent fraction of the spectroscopic parent catalogue.
The DESI selection requires a $W3$ detection, and a complete
audit of individual $W3$ measurement quality was not available
for the matched KiDS and DESI sample. The exploratory comparisons,
for which no multiplicity correction was applied, require
independent confirmation on larger samples with
similar coverage. In particular, control of redshift and apparent
magnitude where possible does not constitute a complete control
for stellar mass, dust content or sample selection.
The models assign galaxies to the available reference classes;
they do not recognise ambiguous morphologies, composite spectra
or other excitation populations as separate categories.
Under these conditions, performance and probability calibration
measured in one survey should not be assumed to apply to another.

%% ======================================================================
\section{The SpecLess software}
\label{sec:specless}

\subsection{Pretrained models and inference}
\label{subsec:specless_interface}

SpecLess is distributed as \texttt{specless-astro} and imported as
\texttt{specless\_astro}. The v0.3.0 release provides
\texttt{MorphologyClassifier} and \texttt{ActivityClassifier},
each initialised for a supported survey. The four XGBoost models
in Table~\ref{tab:packaged} construct the required colours from
prepared catalogue magnitudes and return predictions without
spectroscopic measurements.

Table~\ref{tab:input_summary} summarises the photometric input
specifications. These differ from the ten- and 21-feature
representations used in the reciprocal transfer experiments.

\begin{table*}[t]
\centering
\caption{Inputs to the four pretrained survey-specific classifiers. Colours are constructed from the indicated magnitudes. Photometric system, source measurement type and extinction processing must match the training catalogue; this table records the preparation operations established for the analysis, not a general conversion between survey systems.}
\label{tab:input_summary}
\footnotesize
\begin{tabular}{@{}p{2.1cm}p{2.65cm}p{3.8cm}p{5.8cm}@{}}
\toprule
\textbf{Task} & \textbf{Survey} & \textbf{Input bands and features} & \textbf{Photometric preparation} \\
\midrule
Morphology & SDSS+AllWISE & $ugriz$, $W1,W2$; 7 magnitudes and 21 colours & Galactic-extinction-corrected SDSS optical magnitudes; AllWISE converted from Vega to AB using the stated offsets and infrared extinction corrections. \\
Morphology & KiDS+VIKING & $u,g,r,i_2,Z,Y,J,H,K_s$; 9 magnitudes and 36 colours & KiDS/VIKING GAaP measurements with the adopted uncertainty and quality-flag selection. \\
Activity & SDSS+AllWISE & $ugriz$, $W1,W2,W3$; 8 magnitudes and 28 colours & SDSS optical quality and uncertainty cuts; AllWISE Vega-to-AB conversions and infrared extinction corrections. \\
Activity & DESI Legacy & $g,r,z,W1,W2,W3$; 6 magnitudes, 6 errors and 15 colours & Nanomaggy flux-to-magnitude conversion, propagated uncertainties and band-dependent Galactic-extinction correction. \\
\bottomrule
\end{tabular}
\end{table*}

The input to each model is a pandas DataFrame containing the required photometric bands in the magnitude system and extinction convention of its training catalogue. Survey-specific source selection, cross-matching, photometric quality cuts and flux-to-magnitude conversion are performed upstream. Metadata distributed with each model identify the required bands and derived features; SpecLess does not transform SDSS magnitudes into KiDS GAaP measurements or reconcile AllWISE photometry with DESI forced-WISE measurements. An incorrect survey name or missing photometric column is therefore a data-compatibility issue, not a valid cross-survey inference experiment.

For example, given a prepared DataFrame \texttt{catalogue} containing SDSS-compatible $u,g,r,i,z,W1,W2$ magnitudes, morphology inference is invoked as follows:

The returned table contains the assigned labels and model
probabilities. The \texttt{predict\_proba()} method returns
class probabilities separately. The other models are selected with
\texttt{survey="kids"} for morphology and \texttt{survey="sdss"}
or \texttt{survey="desi"} for activity. Appendix~\ref{app:workflow}
summarises the inference workflow.

\begin{lstlisting}[basicstyle=\ttfamily\footnotesize,frame=single,breaklines=true]
from specless_astro import MorphologyClassifier

classifier = MorphologyClassifier(survey="sdss")
result = classifier.predict(catalogue)
\end{lstlisting}

\subsection{Reproducibility}
\label{subsec:specless_testing}

Inference was tested with the four distributed model/metadata pairs under Python 3.12. Held-out scores were checked against saved predictions or compatible test inputs, and input validation rejected missing and non-finite required fields. The distributed assets reproduce inference, whereas the complete historical catalogue selection and training procedures are not included in the released package. Runtime scaling and cross-version compatibility were not benchmarked for v0.3.0; model files should be loaded from trusted sources and evaluated with compatible software dependencies.

%% ======================================================================

\section{Conclusions}
\label{sec:conclusions}

In this study, we investigated the extent to which galaxy
morphology and spectroscopic activity can be recovered from
broadband photometry across different wavelength ranges and
imaging surveys. Controlled experiments using KiDS+VIKING
and the DESI Legacy Imaging Surveys demonstrate that
classification performance depends on the physical diversity
of galaxy populations and the nature of the available
observational information.

For morphological classification, the contribution of
near-infrared photometry to the recovery of red spiral
galaxies demonstrates that multiband observations can
partially resolve the discrepancy between optical colour
and morphology. For activity classification, the complementary
information provided by optical, near-infrared and
mid-infrared measurements plays an important role in
distinguishing spectroscopically defined galaxy populations.
The dependence of classification performance on H$\alpha$
equivalent width further characterises the extent to which
these populations can be recovered from photometric
measurements.

An important finding of this study is that photometric
measurements from different surveys can lead to different
classifications even when the galaxy identities and reference
labels remain unchanged. This has direct implications for
statistical comparisons of galaxy samples drawn from different
surveys and for estimates of the relative abundances of
galaxy populations.

The open-source SpecLess Python package developed in this
work enables the construction of galaxy samples selected
by morphology and activity from large photometric catalogues
in the absence of spectroscopic observations.
In the context of large-scale observational programmes such
as Rubin and Euclid, this approach provides a basis for
identifying candidate galaxy populations and selecting
targets for follow-up observations in studies of galaxy
evolution.

\section*{Code availability}

The SpecLess source code and pretrained inference models are available at \url{https://github.com/uzayyaydiin/SpecLess} under the BSD-3-Clause licence; the v0.3.0 release (commit \texttt{bf0feadf8d17ada3724e22396b6a9401c93eb0ae}) is archived at \href{https://doi.org/10.5281/zenodo.23516274}{doi:10.5281/zenodo.23516274}. Source survey catalogues are available from SDSS, Galaxy Zoo, ESO KiDS/VIKING, NASA/IPAC AllWISE and the DESI Legacy Imaging Surveys. The software archive does not contain the full derived matched-survey tables or complete historical training records.

\section*{Data availability}

The aggregate numerical results for the original SDSS, KiDS
and DESI transfer experiments are available on Zenodo at
\url{https://doi.org/10.5281/zenodo.23581102}.
The archive includes experiment-level performance metrics,
repeated-split results, feature-representation comparisons,
and transfer analyses. It does not contain the original survey
photometry, individual reference labels, object-level
predictions, or complete historical training records. The
expanded KiDS+VIKING analyses are not included in this archive.
The underlying observational data are accessible through
SDSS, Galaxy Zoo, ESO KiDS/VIKING, NASA/IPAC IRSA AllWISE,
and the DESI Legacy Imaging Surveys.

\section*{Acknowledgements}

This work uses publicly available data from SDSS, KiDS, VIKING, the DESI Legacy Imaging Surveys, AllWISE and Galaxy Zoo. The authors acknowledge the respective survey and archive teams and the Galaxy Zoo volunteers. This research uses KiDS DR5 and VIKING data products from the ESO Science Archive Facility (DOIs: 10.18727/archive/37 and 10.18727/archive/59), produced by the KiDS consortium.

U.A. acknowledges travel support from the Royal Astronomical Society (RAS), which enabled the presentation of this work, and from the European Southern Observatory (ESO), which supported attendance at a VLT data analysis workshop. U.A. also thanks Dr. H. Tu\u{g}\c{c}a \c{S}ener for her encouragement and support.

\appendix
\setcounter{figure}{0}
\renewcommand{\thefigure}{A\arabic{figure}}
\renewcommand{\theHfigure}{appendix.\arabic{figure}}
\section{SpecLess inference workflow}
\label{app:workflow}

\begin{figure*}[t]
    \centering
    \includegraphics[width=0.97\textwidth]{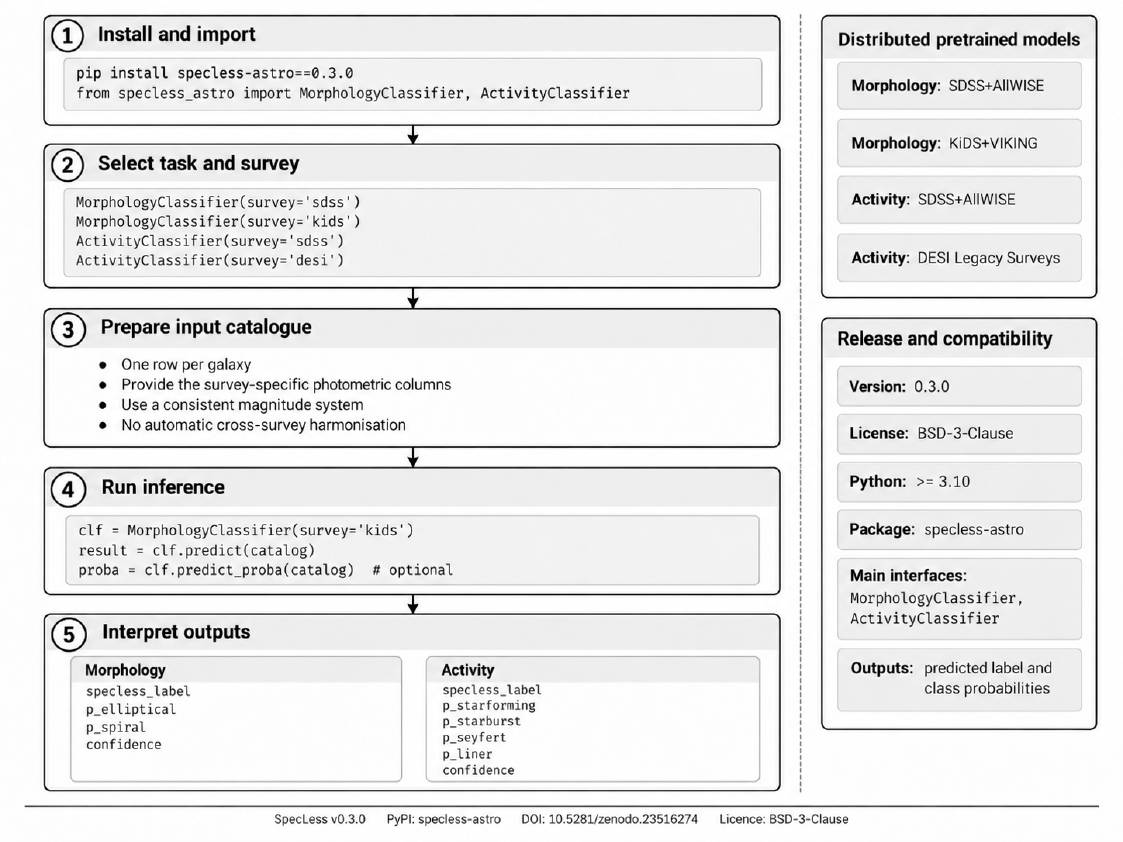}
    \caption{Catalogue-level inference workflow for SpecLess v0.3.0.
    The diagram summarises package installation, selection of a supported
    classification task and survey, preparation of survey-compatible input
    photometry, model inference and interpretation of the returned predictions.
    The side panels identify the four distributed pretrained models and
    release information. Catalogue selection, cross-matching, photometric
    quality cuts, extinction corrections and magnitude-system conversions
    must be carried out before inference as appropriate; the package does not
    automatically harmonise photometry between surveys.
    The explanatory workflow diagram was generated with
    OpenAI ChatGPT.}
    \label{fig:specless_workflow}
\end{figure*}

Figure~\ref{fig:specless_workflow} outlines the use of the four pretrained
classifiers distributed with SpecLess v0.3.0. The user selects a morphology or
activity model for a supported survey and supplies a catalogue containing the
photometric measurements required by that model. The interface reports the
required bands and returns the assigned classes, their probabilities and a
confidence value. The procedure applies to inference with the released model
assets; it does not reproduce the independent training and cross-survey
experiments presented in this paper.

\clearpage

\section*{Declaration of generative AI and AI-assisted
technologies in the manuscript preparation process}

During manuscript preparation, Anthropic Claude was used
to assist with English translation, improve fluency and
reduce translation errors. OpenAI ChatGPT was used to
assist with manuscript revision and generate the explanatory
workflow diagram in Fig.~\ref{fig:specless_workflow}.
The authors reviewed and edited the resulting material
and take full responsibility for its scientific content
and accuracy.
% Before submission, add the verified ChatGPT model/version
% used for Fig. A1 to its caption in accordance with the publisher policy.

\bibliographystyle{elsarticle-harv}
\bibliography{example}

\end{document}